\documentclass[twocolumn,preprintnumbers, amssymb,amsmath,aps,floatfix,prd,nofootinbib,superscriptaddress,showpacs]{revtex4-1}

\usepackage[utf8]{inputenc}

\usepackage{hyperref}
\usepackage{graphicx}

\makeatletter
\let\MYcaption\@makecaption
\makeatother
\usepackage{subcaption}
\makeatletter
\let\@makecaption\MYcaption
\makeatother

\usepackage{bm}
\usepackage{amsmath}
\usepackage{amssymb}

\usepackage{color}
\usepackage{slashed}

\newcommand{\beq}{\begin{eqnarray}}
\newcommand{\eeq}{\end{eqnarray}}

\newcommand{\nn}{\nonumber \\}

\begin{document}

\title{Signature of the  gluon orbital angular momentum}

\author{Shohini Bhattacharya}
\email{sbhattach@bnl.gov  }
\affiliation{Physics Department, Brookhaven National Laboratory, Upton, NY 11973, USA}

\author{Renaud Boussarie}
\email{renaud.boussarie@polytechnique.edu}
\affiliation{CPHT, CNRS, Ecole Polytechnique, Institut Polytechnique de Paris, 91128 Palaiseau, France}

\author{Yoshitaka Hatta}
\email{yhatta@bnl.gov}
\affiliation{Physics Department, Brookhaven National Laboratory, Upton, NY 11973, USA}
\affiliation{RIKEN BNL Research Center, Brookhaven National Laboratory, Upton, NY 11973, USA}

\begin{abstract}
We propose a novel observable for the experimental detection of the gluon orbital angular momentum (OAM)  that constitutes the proton spin sum rule. We consider longitudinal double spin asymmetry in exclusive dijet production in electron-proton scattering and  demonstrate that the $\cos \phi$ azimuthal angle  correlation between the scattered  electron and proton is a sensitive probe of the gluon OAM at small-$x$ and its interplay with the gluon helicity. We also present a numerical estimate of  the cross section for the kinematics of the Electron-Ion Collider.
\end{abstract}

\maketitle

{\it 1. Introduction}---After more than 20 years of operation, the  Relativistic Heavy Ion Collider (RHIC)  spin program at Brookhaven National Laboratory has revealed that the gluon helicity contribution $\Delta G$ to the proton spin sum rule
\beq
\frac{1}{2}=\frac{1}{2}\Delta \Sigma + \Delta G + L_q + L_g,
\eeq
is nonvanishing and likely sizable  \cite{STAR:2014wox,deFlorian:2014yva,Nocera:2014gqa,Ethier:2017zbq,STAR:2021mqa}. Together with the known quark helicity contribution $\Delta\Sigma\sim 0.3$, the result  indicates that parton helicities account for a significant fraction of the proton spin. Yet, there still remain huge uncertainties about the small-$x$ contribution to $\Delta G=\int_0^1 dx \Delta G(x)$ defined as the first moment of the polarized gluon distribution function. Resolving this issue is one of the major goals of the future Electron-Ion Collider (EIC) \cite{AbdulKhalek:2021gbh}.  

Another obvious goal of the EIC is to measure the orbital angular momentum (OAM) of quarks and gluons $L_{q,g}$. However,  progress in this direction is relatively  slow although there have been continuing theory efforts (for recent works, see, e.g., \cite{Boussarie:2019icw,Kovchegov:2019rrz,Engelhardt:2020qtg,Guo:2021aik}). Currently, there does not seem to be a consensus in the community about which observables need to be measured at the EIC in order to constrain $L_{q,g}$. This is so even after the first wave of proposals for experimental observables  appeared several years ago  \cite{Ji:2016jgn,Hatta:2016aoc,Bhattacharya:2017bvs,Bhattacharya:2018lgm,Guo:2021aik} (see also an earlier attempt \cite{Courtoy:2013oaa}).  These works  exploit the known connection \cite{Lorce:2011kd,Hatta:2011ku,Lorce:2011ni} between parton OAMs and the Wigner distributions \cite{Belitsky:2003nz}, or equivalently, the generalized transverse momentum dependent distributions  (GTMDs).   
However, the required processes typically involve very  exclusive final states which are challenging to measure. Besides,  observables are often related to GTMDs via complicated multi-dimensional convolutions even at the leading order. Clearly, more theoretical efforts are needed to increase the accuracy of predictions or devise new observables better suited for the purpose.       

With this in mind, we take a fresh look at the process considered in  Refs.~\cite{Ji:2016jgn,Hatta:2016aoc}. 
There, the authors have proposed to measure longitudinal single spin asymmetry (SSA) in exclusive dijet production in electron-proton collisions $ep \to \gamma^*p \to jjp'$  where the incoming proton is longitudinally polarized. It has been shown that the following angular-dependent part of the cross section
\beq
d \sigma^{h_p}\sim h_p \sin(\phi_{q_\perp}-\phi_{\Delta_\perp})(z-\bar{z})  \mathfrak{Im} (A_2 A^*_3), \label{ssa}
\eeq
is an experimental probe of the gluon OAM. $h_p=\pm 1$ is the proton helicity and $z$ ($\bar{z}=1-z$) is the momentum fraction of the virtual photon carried by the quark (antiquark)  jet.   $\phi_{q_\perp}$ is the azimuthal angle of the relative transverse momentum of the two jets  $q_\perp=q_{1\perp}-q_{2\perp}$ in a frame in which the proton and virtual photon are collinear, and  $\phi_{\Delta_\perp}$ is that of the scattered proton.  $A_2$ and $A_3$ are certain twist-two and twist-three amplitudes, respectively, and the latter is sensitive to the gluon OAM. 
As is familiar in the context of transverse single spin asymmetry, one takes the imaginary part of their  interference terms.

In this paper, we propose a new observable for the gluon OAM by  implementing  two major changes in the above proposal. First, we consider {\it double} spin asymmetry (DSA) in dijet production  $e p \to e'jjp'$ where both the electron and incoming proton are longitudinally polarized.   The outgoing lepton must be  tagged and its azimuthal angle  $\phi_{l_\perp}$ measured. Electron polarization brings in an extra factor of $i$ to the cross section, so this time one takes the real part of the interference terms. The formula we shall arrive at is 
\beq
d \sigma^{h_ph_l} \sim h_p h_l\cos (\phi_{l_\perp}-\phi_{\Delta_\perp}) \mathfrak{Re}(A'_2A'^*_3) ,\label{dsa}
\eeq
where $h_l=\pm 1$ is the electron helicity. Experimentally,  the term (\ref{dsa}) can be isolated by forming the linear combination $d\sigma^{++}-d\sigma^{+-}-d\sigma^{-+}+d\sigma^{--}$. Unlike in (\ref{ssa}), the prefactor does not vanish for symmetric jet configurations ($z=1/2$), a fact which will turn out to be important. We shall argue that DSA (\ref{dsa}) is more advantageous than SSA (\ref{ssa}) from both theoretical and practical points of view.  Second, we point out that there is another contribution to the asymmetry coming from the gluon helicity generalized parton distribution (GPD). Such a contribution was overlooked in \cite{Ji:2016jgn,Hatta:2016aoc}, but is parameterically  as important as that from the OAM. We shall perform the leading order calculation of both contributions and numerically evaluate them. The result   demonstrates that   DSA in dijet production is a unique observable which allows us to directly probe into the gluon OAM $L_g$ and its interplay with the gluon helicity $\Delta G$. 
\\

{\it 2. Orbital angular momentum and GTMDs}---Let us first quickly recapitulate the connection between GTMDs and parton OAMs. Following \cite{Meissner:2009ww,Lorce:2013pza}, we parameterize the leading-twist gluon GTMDs as 
\beq
&&xf_g(x,\xi,\widetilde{k}_\perp,\widetilde{\Delta}_\perp) \nn 
&&=\int \!\frac{d^3z}{(2\pi)^3P^+} e^{ixP^+z^-\!-i\widetilde{k}_\perp\cdot z_\perp} \langle p'|F_a^{+i}(-z/2)F_a^{+i}(z/2)|p\rangle \nn 
&&= \frac{1}{2M}\bar{u}(p')\biggl[ F_{1,1}+i\frac{\sigma^{j+}}{P^+}(\widetilde{k}_\perp^j F_{1,2}+\widetilde{\Delta}_\perp^j F_{1,3}) \nn
&& \qquad  + i\frac{\sigma^{ij}\widetilde{k}_\perp^i\widetilde{\Delta}_\perp^j}{M^2}F_{1,4}\biggr]u(p), \label{gtmd1}
\eeq  
where $P^\mu = \frac{p+p'}{2}$, $\Delta^\mu=p'^\mu- p^\mu$ and $\xi=(p^+-p'^+)/2P^+$. $i,j=1,2$ are two-dimensional vector indices.  All the GTMDs are a function of $x,\xi,\widetilde{k}_\perp^2,\widetilde{\Delta}_\perp^2$ and $\widetilde{k}_\perp\cdot \widetilde{\Delta}_\perp$. 
The usual GPDs are obtained by integrating over $\widetilde{k}_\perp$,
\beq
\int d^2k_\perp xf_g &=& \!\frac{1}{2P^+} \bar{u}(p')\left(\! H_g\gamma^++\! E_g\frac{i\sigma^{+\nu}\Delta_\nu}{2M}\right)\! u(p),
\eeq
normalized as $H_g(x)=xG(x)$ in the forward limit. In the following we shall encounter the integrals 
\beq
xL_g (x,\xi )&\equiv&  -\int d^2\widetilde{k}_\perp \frac{\widetilde{k}_\perp^2}{M^2}F_{1,4}(x,\xi,\widetilde{\Delta}_\perp=0) ,\label{lgx}  \\
O(x,\xi )&\equiv&  \int d^2\widetilde{k}_\perp \frac{\widetilde{k}_\perp^2}{M^2}F_{1,2}(x,\xi,\widetilde{\Delta}_\perp=0). \label{o}
\eeq
In the limit $\xi\to 0$, $L_g(x)$ is the parton distribution function of  the gluon OAM \cite{Hatta:2012cs} normalized as  $\int_0^1dx L_g(x)=L_g$. The imaginary part of $F_{1,2}$ is called the spin-dependent Odderon \cite{Zhou:2013gsa} and its $k_\perp$-moment is related to the three-gluon correlator relevant to transverse single spin asymmetry. The real part of $F_{1,2}$ is proportional to $\xi$, but otherwise unconstrained.  
In (\ref{gtmd1}), the GTMDs are  defined in the `symmetric' frame where $P_\perp=0$ so that $\widetilde{p}'_\perp = \widetilde{\Delta}_\perp/2 = -\widetilde{p}_\perp$. The advantage of this frame is that one can exploit  $PT$  (parity \& time-reversal) symmetry to constrain the dependence of GTMDs on variables.   
However, this frame is inconvenient and practically not used   when describing actual experimental  processes. We shall instead work in the so-called hadron frame where the incoming virtual photon and proton are collinear along the $x^3$ direction, namely, $p_\perp=0$.  
The two frames are related by the so-called  transverse boost, a Lorentz transformation that leaves invariant the plus component of a four-vector 
$V^+ = \widetilde{V}^+$,  
$V_\perp = \widetilde{V}_\perp + C_\perp \widetilde{V}^+$, 
$V^- = \widetilde{V}^- + C_\perp \cdot \widetilde{V}_\perp +\frac{C_\perp^2}{2}\widetilde{V}^+$.  Applying this transformation to the matrix element (\ref{gtmd1}) with $C_\perp=\widetilde{\Delta}_\perp/(2p^+)$, we see that if one considers a scattering process in the hadron frame where $p_\perp=0$, transverse momentum transfer $p'_\perp=\Delta_\perp$ and $t$-channel gluons with transverse momentum $k_\perp$, the GTMDs $F_{1,n}(x,\xi,\widetilde{k}_\perp,\widetilde{\Delta}_\perp)$ should be evaluated at 
\beq
 \widetilde{k}_\perp 
 = k_\perp -\frac{x}{2}\Delta_\perp, \quad  \widetilde{\Delta}_\perp= (1+\xi)\Delta_\perp. \label{difference}
\eeq
Many of the previous phenomenological applications of GTMDs have adopted  the small-$x$ kinematics $x\ll 1$ which also implies $\xi\ll 1$. In such cases, the difference (\ref{difference}) is negligible to first approximation. \\
\begin{figure}
  \includegraphics[width=1\linewidth]{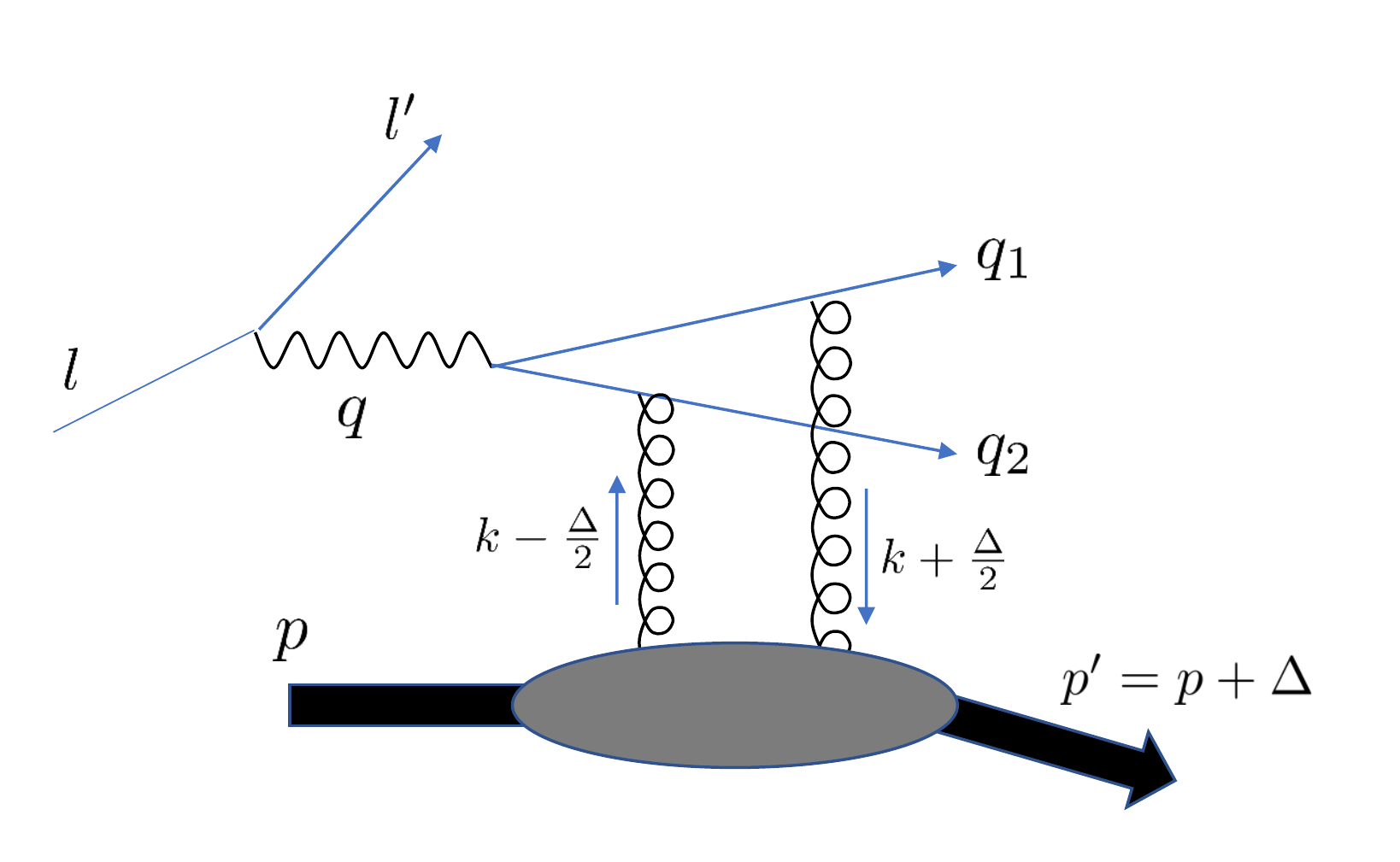}
\caption{Exclusive dijet production in electron-proton scattering.}
\label{1}
\end{figure} 

{\it 3 Double spin asymmetry in diffractive dijet production}---We consider exclusive dijet production in electron-proton scattering depicted in Fig.~\ref{1}. This process has attracted a lot of attention in the literature in different contexts \cite{Bartels:1996ne,Braun:2005rg,Altinoluk:2015dpi,Hatta:2016dxp,Boussarie:2016ogo,Ji:2016jgn,Hatta:2016aoc,Hagiwara:2017fye,Mantysaari:2019csc,Salazar:2019ncp,Boer:2021upt}. However, longitudinal double spin asymmetry has not been studied to our knowledge.  
In the hadron frame,  the longitudinally polarized proton moves fast in the $+x^3$ direction and the virtual photon with virtuality $q^2=-Q^2$ in the $-x^3$ direction. Two jets in the final state have longitudinal momentum fractions (of the photon) $z=p\cdot q_1/p\cdot q$ and $\bar{z}=1-z$, and transverse momenta $q_\perp-z\Delta_\perp$ and $-q_\perp-\bar{z}\Delta_\perp$, respectively, where $\Delta_\perp$ is the transverse momentum of the recoiling proton. $q_\perp$ is related to skewness $\xi$ and the $\gamma^*p$ center-of-mass energy $W^2=(p+q)^2$ as 
\beq
\xi = \frac{q_\perp^2+z\bar{z}Q^2}{-q_\perp^2+z\bar{z}(Q^2+2W^2)}. \label{xidef}
\eeq
The momenta of the incoming lepton is parameterized as $l^\mu= (l^+,l^-,l_\perp)=(\frac{Q(1-y)}{\sqrt{2}y}, \frac{Q}{\sqrt{2}y}, \frac{Q\sqrt{1-y}}{y}n_\perp)$  
where $y=p\cdot q/p\cdot l$ as usual and $n_\perp=(\cos \phi_{l_\perp},\sin\phi_{l_\perp})$ is a unit vector in the transverse plane.  
The spin-dependent part of the  lepton tensor is 
$L^{\mu\nu} \sim  -2i\epsilon^{\mu\nu\rho\sigma}s_\rho q_\sigma$. 
For a longitudinally polarized lepton, $s_\rho =h_l l_\rho$ where $h_l=\pm 1$ is the helicity. In order to be sensitive to the azimuthal angle   of the lepton plane $\phi_{l_\perp}$, the index $\rho$ has to be transverse. Since $\sigma=\pm$ is longitudinal, one of $\mu,\nu$ must be longitudinal and the other transverse. Namely, we should look at the interference effect between the longitudinal $A_L$ and transverse $A_T^\lambda$ $(\lambda=1,2$) virtual  photon  amplitudes
\beq
A_L=A_L^2+A_L^3, \quad A^\lambda_T 
= \epsilon^{\lambda i} (A_{T}^{2i}+A_{T}^{3i}).
\eeq
The twist-2 part $A^2_{L/T}$ is proportional to gluon GPDs and  has been calculated in \cite{Braun:2005rg} in the two-gluon exchange approximation (see Fig.~\ref{1}).  The twist-3 part $A^3_{L/T}$ involves GTMDs and retains one factor of  $t$-channel gluon transverse momentum $k_\perp$. It has been calculated in \cite{Ji:2016jgn} and here we reproduce the result 
\begin{align}
A^{3}_{T} & = - \dfrac{i g^{2}_s e_{em} e_q}{N_c} \dfrac{2(\overline{z}-z)}{(q^{2}_\perp + \mu^{2})^{2}}  \bar{u}(q_1) \epsilon_\perp \cdot \gamma_\perp v(q_2) \nn 
&\quad \times\int dx \dfrac{x}{ (x^2 - \xi^2 +i \xi\varepsilon)^{2}}  \bigg ( 2\xi + \dfrac{(2\xi)^{3}(1-2\beta)}{(x^2-\xi^2 + i\xi \varepsilon)}\bigg )\nn 
&\quad \times \int d^{2}k_\perp q_\perp \cdot k_\perp \, x f_{g}(x,\xi,k_\perp,\Delta_\perp) \nonumber \\[0.2cm]
& - \dfrac{i g^{2}_s e_{em} e_q}{N_c} \dfrac{2 (2\xi)^{2} z \overline{z} W}{(q^{2}_\perp + \mu^{2})^{2}}  \bar{u}(q_1) \gamma^- v(q_2) \nn
& \quad \times \int dx \dfrac{x}{(x^2 - \xi^2 +i\xi \varepsilon)^{2} } \nn
& \quad \times \int d^{2}k_\perp \epsilon_\perp \cdot k_\perp \, x f_{g}(x,\xi,k_\perp,\Delta_\perp), \label{a3t} \\[0.3cm] 
A^{3}_{L} & = \dfrac{i g^{2}_s e_{em} e_q}{N_c} \dfrac{16 \xi^{2}(\overline{z}-z)z\overline{z}Q W}{(q^{2}_\perp + \mu^{2})^3}  \bar{u}(q_1) \gamma^- v(q_2)\nn
&\quad \times \int dx \dfrac{x}{(x^2 - \xi^2 +i\xi  \varepsilon)^2 } 
\bigg ( 1 + \dfrac{8 \xi^{2}(1-\beta)}{(x^2-\xi^2+ i\xi \varepsilon) } \bigg ) \nn
&\quad  \times \int d^{2}k_\perp \, q_\perp \cdot k_\perp \, x f_g (x,\xi, k_\perp, \Delta_\perp ) , \label{a3l}
\end{align}
where $\mu^2=z\bar{z}Q^2$ and $\beta=\frac{\mu^2}{q_\perp^2+\mu^2}$. 
The $k_\perp$-weighted integrals of $f_g$ lead to the  moments (\ref{lgx}) and (\ref{o}). Importantly, in both the longitudinal and transverse amplitudes, the $x$-integral contains a third pole at $x=\pm \xi$. Such poles often imply the breakdown of collinear factorization due to diverging $x$-integrals \cite{Cui:2018jha}. (Gluon GPDs may contain terms proportional to  $\theta(\xi-x)(x^2-\xi^2)^2$ which are not integrable if there is a third pole.) Fortunately,  these potentially dangerous terms can be dropped by setting $z=1/2$, after which $A_L^3=0$ and only a second pole remains in $A_T^3$. Note that, if one considers SSA \cite{Ji:2016jgn}, one cannot set $z=1/2$ because the asymmetry (\ref{ssa}) vanishes at this point.  After integrating over the jet azimuthal angle $\phi_{q_\perp}$, we obtain the following contribution to DSA at $z=1/2$   
\begin{align}
&\dfrac{d\sigma}{dy dQ^{2} d \phi_{l_\perp}dz dq_\perp^2 d^{2}\Delta_\perp }   \nn
& =\dfrac{\alpha_{em}y}{2^{11}\pi^7Q^4} 
\dfrac{\int d\phi_{q_\perp} L^{\mu \nu} A^{*}_{\mu} A_{\nu}}{(W^{2}+Q^{2})(W^{2}-M_J^{2})z\overline{z}}, \label{cross}
\end{align}
where $M_J^2=q_\perp^2/(z\bar{z})=4q_\perp^2$ is the invariant mass of dijet and 
\begin{align}
& \int d\phi_{q_\perp} L^{\mu\nu}A^*_\mu A_\nu 
\nn & = -\dfrac{2^{10} \pi^{4}}{N_c}h_lh_p \alpha^{2}_s \alpha_{em} e^{2}_q  \frac{ (1+\xi) \xi  Q^{2}}{(q_\perp^2+\mu^2)^2}|l_\perp|| \Delta_\perp|   \nn 
& \  \times \mathfrak{Re}\Bigg [   \biggl\{ {\cal H}^{(1)*}_g - \dfrac{\xi^{2}}{1-\xi^{2}}{\cal E}^{(1)*}_g   \nonumber \\[0.2cm]
& \qquad  +   \dfrac{4q^{2}_\perp}{q^{2}_\perp +\mu^{2}}  \bigg ( {\cal H}^{(2)*}_g - \dfrac{\xi^{2}}{1-\xi^{2}}{\cal E}^{(2)*}_g \bigg )\biggr\} {\cal L}_g  
\nonumber \\[0.2cm]
& \quad  + \left( {\cal E}^{(1)*}_g 
+  \dfrac{4q^{2}_\perp}{q^{2}_\perp +\mu^{2}}   {\cal E}^{(2)*}_g \right) \frac{{\cal O}}{2} \Bigg ]\textrm{cos}(\phi_{l_\perp}  - \phi_{\Delta_\perp}).
\label{main}
\end{align}
The details of the calculation, including the case $z\neq 1/2$, will be presented elsewhere \cite{prep}. 
The various `Compton form factors' are defined as 
\beq
{\cal H}^{(1)}_g(\xi)&=&\int^1_{-1} dx \frac{ H_g(x,\xi)}{(x-\xi+i\epsilon) (x+\xi-i\epsilon)} , \label{h1} \\ 
{\cal H}^{(2)}_g(\xi)&=&\int^1_{-1} dx  \frac{\xi^2 H_g(x,\xi) }{(x-\xi+i\epsilon)^2 (x+\xi-i\epsilon)^2}, \label{h2} \\
{\cal L}_g(\xi) &=&\int^1_{-1} dx \frac{x^2L_g(x,\xi)}{(x-\xi+i\epsilon)^2 (x+\xi-i\epsilon)^2} \label{l1} ,\\
{\cal O}(\xi) &=&\int^1_{-1} dx \frac{xO(x,\xi)}{(x-\xi+i\epsilon)^2 (x+\xi-i\epsilon)^2},
\eeq
and ${\cal E}_g^{(1,2)}(\xi)$ is defined from $E_g(x,\xi)$ similarly to ${\cal H}_g^{(1,2)}(\xi)$. 
Assuming $|H_g|\gg |E_g|$, we see that the cross section is directly proportional to the Compton form factor of the gluon OAM ${\cal L}_g$. The characteristic  correlation $k_\perp \times \Delta_\perp\sim \sin (\phi_{k_\perp}-\phi_{\Delta_\perp})$ of OAM manifests itself as a cosine  correlation between the outgoing electron and proton angles. A similar transfer of angular correlations  has been noticed   in \cite{Zhou:2016rnt} for the $\cos 2(\phi_{k_\perp}-\phi_{\Delta_\perp})$ dependence of the elliptic gluon GTMD \cite{Hatta:2016dxp,Hagiwara:2021xkf}. Away from the point $z=1/2$, there are corrections proportional to $(z-1/2)^2$, but collinear factorization is suspect for them as already mentioned. Instead, one should use the $k_\perp$-factorization approach to calculate the   corrections, although their connection to the OAM is less clear. 
\\

{\it 4. DSA from the gluon helicity}---Next we discuss another source of DSA from the gluon helicity GPDs 
\beq
\epsilon_{ij}\int \frac{dz^-}{\pi } e^{ixP^+z^-}\langle p'|F_a^{+i}(-z/2)F_a^{+j}(z/2)|p\rangle  \nn
= \bar{u}(p')\left(i\tilde{H}_g\gamma_5\gamma^+-i \tilde{E}_g\frac{\gamma_5\Delta^+}{2M}\right)u(p),
\label{e:GPD_para}
\eeq
where $\tilde{H}_g(x)=x\Delta G(x)$ in the forward limit.   
This originates from the interference between the unpolarized and polarized gluon GPDs in the amplitude and the complex-conjugate amplitude.\footnote{We mention in passing that if one starts out with the GTMD version of Eq.~(\ref{e:GPD_para}), then there will  be another contribution to the asymmetry proportional to  certain gluon helicity GTMDs and the gluon helicity GPD. However, we expect this contribution to be orders of magnitude smaller than the one we discuss here, because the helicity GPD would be much smaller than the unpolarized GPD in the kinematics we consider.
We plan to explain these subtleties in a follow-up work \cite{prep}.}
An entirely analogous contribution should be added to the result for  SSA in  \cite{Ji:2016jgn,Hatta:2016aoc}. The $\cos (\phi_{l_\perp}-\phi_{\Delta_\perp})$ asymmetry due to  this mechanism  
is actually known in the context of Deeply Virtual Compton Scattering (DVCS) \cite{Belitsky:2000gz}. Unlike in DVCS, in dijet production there is no contamination from the Bethe-Heitler process. The asymmetry can be calculated purely within the GPD framework by setting  $k_\perp=0$ but keeping one factor of $\Delta_\perp$ in the hard part. 
In general, the cross section contains integrals with a third pole such as 
\beq
\int dx \frac{H_g(x,\xi)}{(x^2-\xi^2+i\xi \epsilon)^3}, \quad \int dx \frac{x\tilde{H}_g(x,\xi)}{(x^2-\xi^2+i\xi \epsilon)^3}. \label{third}
\eeq
Remarkably, however, these factorization-breaking terms all vanish at $z=1/2$ and we find \cite{prep}
\beq
&& \int d\phi_{q_\perp} L^{\mu\nu}A_\mu A_\nu = \frac{2^{10}\pi^4}{N_c}h_lh_p\alpha_s^2\alpha_{em}e_q^2 \frac{(1-\xi^2)\xi Q^2}{(q_\perp^2+\mu^2)^2}\nn
&&\quad \times \mathfrak{Re}\left[\left({\cal H}_g^{(1)*} - \frac{\xi^2}{1-\xi^2}{\cal E}_g^{(1)*}\right)\left(\tilde{\cal H}_g^{(2)} - \frac{\xi^2}{1-\xi^2}\tilde{\cal E}_g^{(2)}\right) \right] \nn 
&& \quad \times |l_\perp ||\Delta_\perp| \cos (\phi_{l_\perp}-\phi_{\Delta_\perp}), \label{main2}
\eeq
with 
\beq
\tilde{\cal H}_g^{(2)}(\xi) = \int dx \frac{x \tilde{H}_g(x,\xi)}{(x^2-\xi^2+i\xi\epsilon)^2}, \nn \tilde{\cal E}_g^{(2)}(\xi) = \int dx \frac{x \tilde{E}_g(x,\xi)}{(x^2-\xi^2+i\xi\epsilon)^2}.
\eeq

In the following, we shall be mainly interested in the small-$\xi$ region $\xi \lesssim 10^{-3}$. In this region ${\cal H}_g^{(1,2)}(\xi)$ are dominated by the imaginary part, and one can show that  $\mathfrak{Im}{\cal H}_g^{(1)} \approx -2\mathfrak{Im}{\cal H}_g^{(2)}$.\footnote{More precisely, one can show that 
\beq
\mathfrak{Im} {\cal H}^{(2)}_g(\xi) = - \dfrac{\pi}{2} \dfrac{d}{dx}{H_g(x,\xi)} \big |_{x=\xi} + \dfrac{\pi}{2\xi} H_g (\xi,\xi) . \nonumber
\eeq
In the limit $\xi\ll 1$, the second term dominates. 
}
Combining (\ref{main2}) with (\ref{main}) and neglecting ${\cal E}_g$ and $\tilde{\cal E}_g$, we find that the asymmetry is roughly proportional to the combination 
\beq
{\cal H}^{(1)*}_g \left(\tilde{\cal H}_g^{(2)} +\frac{q_\perp^2-\mu^2}{q_\perp^2+\mu^2} {\cal L}_g\right). \label{cancel}
\eeq
Depending on the sign of $q_\perp^2-\mu^2=q_\perp^2-Q^2/4$, the helicity and OAM contributions interfere positively or negatively. Note that $\tilde{\cal H}_g^{(2)}\pm {\cal L}_g \sim \Delta G(x)\pm L_g(x)$, and $\Delta G(x)\approx -L_g(x)$ at small-$x$   \cite{Hatta:2016aoc,Hatta:2018itc,More:2017zqp,Boussarie:2019icw}  (see however, \cite{Kovchegov:2019rrz}). Thus, the two contributions have the same sign  when $q_\perp^2 < Q^2/4$ but tend to cancel each other when $q_\perp^2>Q^2/4$.  By varying $Q^2$, we should be able to see this very interesting interplay between the helicity and the OAM.  \\

\begin{figure*}[t]
\begin{minipage}{0.3\hsize}
\includegraphics[scale=0.4]{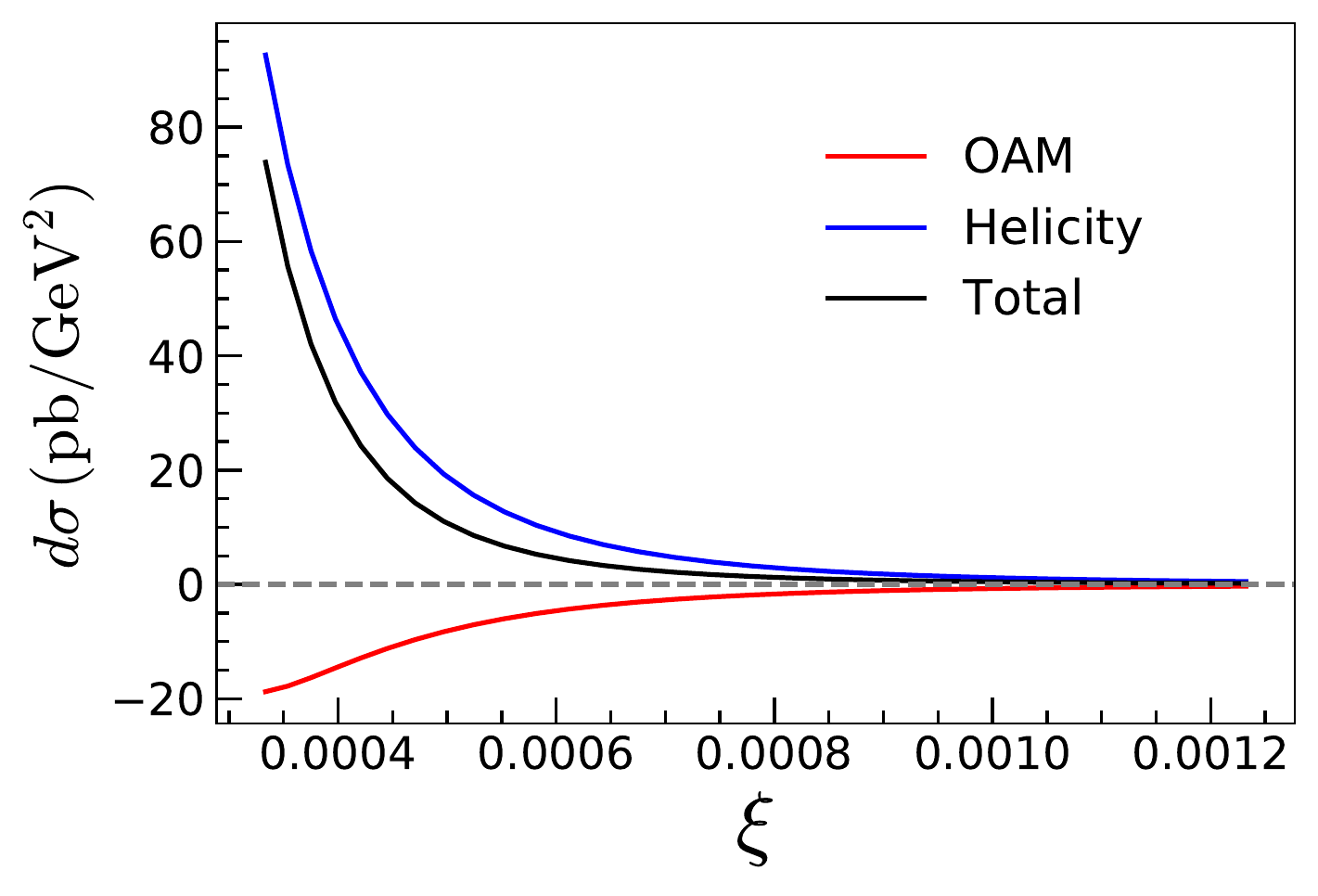}
\end{minipage}
\hfill
\begin{minipage}{0.3\hsize}
\hspace*{-0.2cm}\includegraphics[scale=0.4]{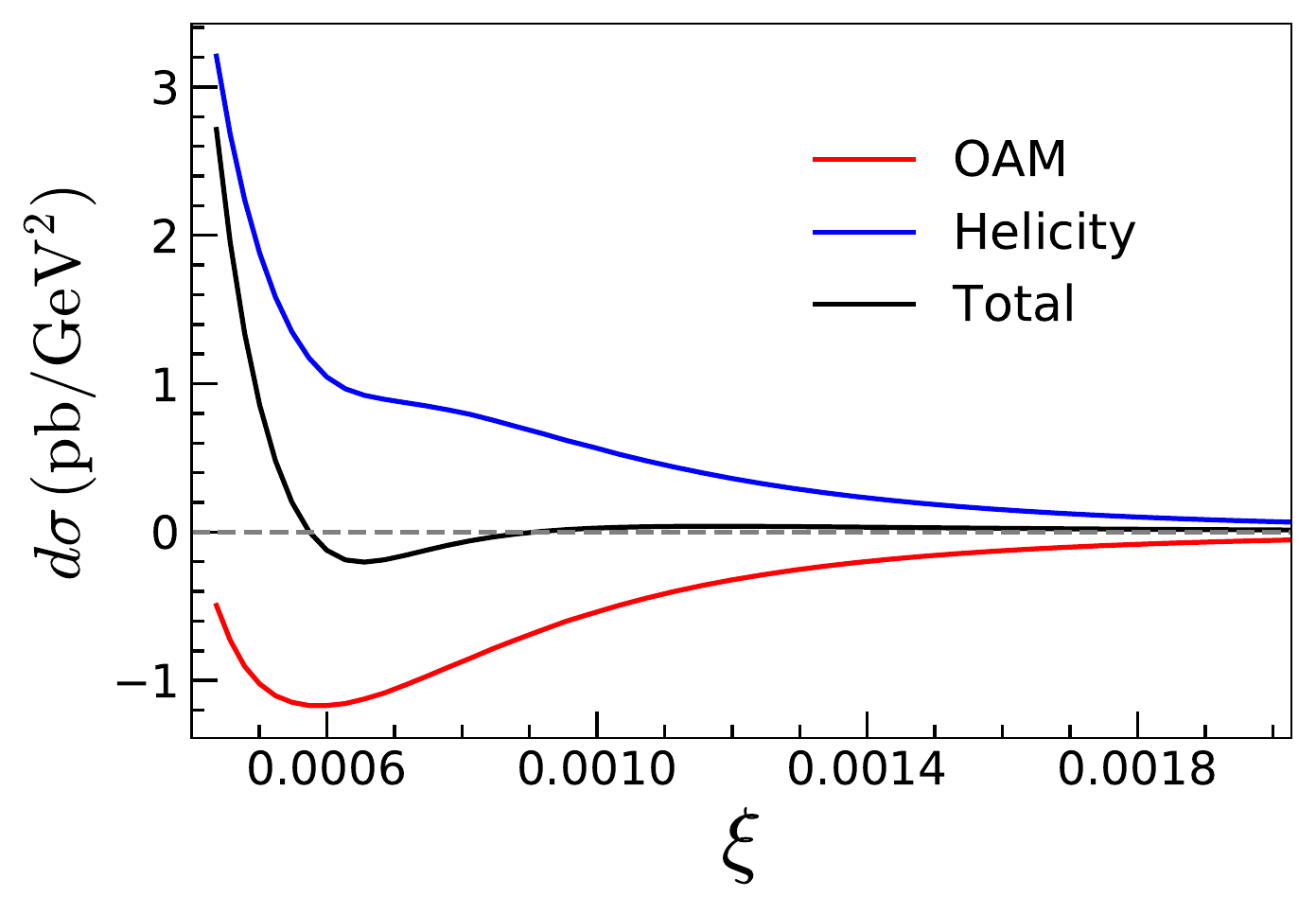}
\end{minipage}
\hfill
\begin{minipage}{0.3\hsize}
\hspace*{-0.6cm}\includegraphics[scale=0.4]{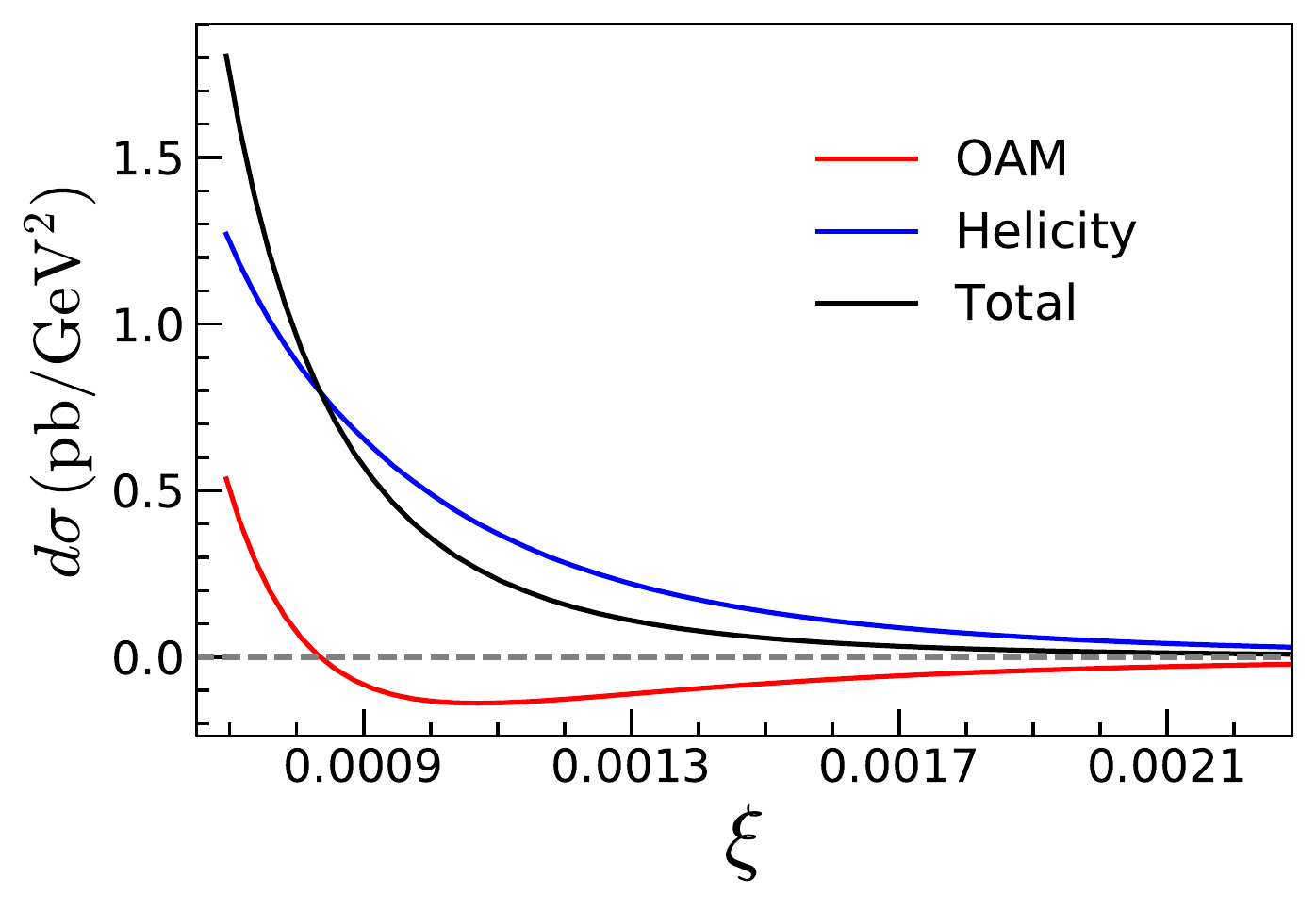}
\end{minipage}
\caption{The DSA  part of the differential cross section (\ref{cross}) at $Q^2=2.7$ GeV$^2$ (left), $Q^2=4.8$ GeV$^2$ (middle) and $Q^2=10$ GeV$^2$ (right).  The labels `OAM' and `Helicity' refer to the two contributions (\ref{main}) and (\ref{main2}), respectively. }
\label{2}
\end{figure*}

{\it 5. Numerical results}---We now present a numerical estimate of the cross section. We neglect $E_g,\tilde{E}_g$ altogether.
$H_g(x,\xi)$, $\tilde{H}_g(x,\xi)$ and $xL_g(x,\xi)$ are reconstructed  from their PDF counterparts  $xG(x)$, $x\Delta G(x)$ and $xL_g(x)$, respectively,  using the method of double distributions  \cite{Radyushkin:1998es,Radyushkin:2000uy}. We use the JAM~\cite{Sato:2019yez,Ethier:2017zbq} gluon PDFs $xG(x)$ and $x\Delta G(x)$  as inputs.     
As for $xL_g(x)$, we employ the  Wandzura-Wilczek (WW) approximation  \cite{Hatta:2012cs} 
\beq
L_g(x) \approx x\int^1_x \frac{dx'}{x'^2}x'G(x') - 2x \int_x^1 \frac{dx'}{x'^2}\Delta G(x') ,
\eeq
although we are eventually interested in constraining the genuine twist-three part neglected in this approximation. 
We integrate over  $\Delta_\perp$ assuming a Gaussian form factor $e^{-b\Delta_\perp^2}$ with $b=5$ GeV$^{-2}$ \cite{Braun:2005rg} and change variables $q_\perp \to \xi$ according to  (\ref{xidef}). The other parameters are fixed as $h_p h_l=1$,  $\sqrt{s_{ep}}=120$ GeV and $y=0.7$.  The resulting cross section (only the DSA part)  
\beq
\frac{d \sigma}{dydQ^2dz d\xi d\delta \phi }, \label{cross}
\eeq
is shown in Fig.~\ref{2} at $\delta\phi=\phi_{l_\perp}-\phi_{\Delta_\perp}=0$ for three different values of $Q^2$ (2.7 GeV$^2$,  4.8  GeV$^2$ and 10 GeV$^2$). The plots correspond to $1<q_\perp<3$ GeV ($1< q_\perp <2.35$ GeV in the $Q^2=2.7$ GeV$^2$ case). Typical jet rapidities  in the laboratory frame  are $-2<\eta <-1$ at the top EIC energy $E_p\sim 250$ GeV.  We see that  the OAM (\ref{main}) and helicity (\ref{main2}) contributions are comparable in magnitude, though the latter tends to be larger because of the cancellation between ${\cal H}^{(1)}_g$ and ${\cal H}_g^{(2)}$. As a result of this cancellation, we observe a clear sign change of the OAM contribution with increasing $Q^2$, see (\ref{cancel}).    
It should be mentioned that  there are large uncertainties in our prediction even in the helicity part because currently $\Delta G(x)$ is poorly constrained, including even the sign, especially in the small-$x$ region but also in the large-$x$ region. (See a recent discussion \cite{Zhou:2022wzm} on this point.)  Besides, nothing is known about $L_g(x)$ experimentally at the moment, and our model for ${\cal L}_g$    involves key assumptions (the WW approximation and the use of the double distribution technique) whose validity needs to be investigated.   The above result should thus be regarded as an exploratory study to be significantly improved in future.  
Nonetheless, our calculation adequately demonstrates the feasibility of accessing the OAM from DSA. Ultimately, ${\cal L}_g$ can be extracted from future experimental data, and for this purpose an  accurate determination of $xG(x)$ and $x\Delta G(x)$ down to $x\sim 10^{-3}$ is crucial. 
\\

{\it 6. Conclusions}---We have proposed DSA in exclusive dijet production as a novel observable for the gluon OAM $L_g$ that can be measured at the EIC. Compared to SSA (\ref{ssa}) previously suggested in  \cite{Ji:2016jgn,Hatta:2016aoc}, it has a number of advantages. Most importantly, the third poles at $x=\pm \xi$ in (\ref{a3t}), (\ref{a3l}) and (\ref{third}) which are potentially dangerous for QCD factorization can be eliminated by setting $z=1/2$, but this is not possible in SSA. In practice,  measurements are done in some  window in $z$. We expect that the cross section $d\sigma/dz$ varies mildly around $z\sim 1/2$, but this  needs to be substantiated in future investigations. Second, unlike the jet angle $\phi_{q_\perp}$, the electron angle $\phi_{l_\perp}$ is not affected by final state QCD radiations.  The former is integrated over in DSA, and this greatly simplifies the cross section formula  without losing sensitivity to the OAM. We thus expect that DSA is more robust against higher order QCD corrections to this process \cite{Boussarie:2016ogo}. Furthermore, in the limit $x\approx \xi\to 0$, ${\cal H}_g^{(1,2)}$ are dominantly imaginary, and the extraction of the imaginary part in (\ref{ssa}) turned out to be a delicate problem within the  effective theory of high energy QCD \cite{Hatta:2016aoc}. For DSA, such a concern is simply absent. (We however note that in the present GPD-like approach, the real and imaginary parts of ${\cal L}_g$ are comparable in magnitude.)

The present calculation can  be straightforwardly extended to the quark exchange channel important in the low-energy (low-$W$, high-$\xi$) region. 
We expect an additional contribution  proportional to the product of the quark GPD and the quark OAM $\sim {\cal H}^*_q{\cal L}_q$. This will be a nice addition to the finding in \cite{Bhattacharya:2017bvs} which is so far the only observable known to be sensitive to $L_q$. \\

{\it Acknowledgements}---We thank Feng Yuan and  Yong Zhao for explaining to us the results in  \cite{Ji:2016jgn} and for discussion.
  S.~B. and Y.~H. were supported by the U.S. Department of Energy under Contract No. DE-SC0012704, and also by  Laboratory Directed Research and Development (LDRD) funds from Brookhaven Science Associates.  S.~B. has also been supported by the U.S. Department of Energy, Office of Science, Office of Nuclear Physics and Office of Advanced Scientific Computing Research within the framework of Scientific Discovery through Advance Computing (SciDAC) award Computing the Properties of Matter with Leadership Computing Resources.


\end{document}